\documentclass[twocolumn,prb,floats,superscriptaddress, aps]{revtex4-2}

\newcommand{\tw}[1]{\ensuremath{t_{w_{#1}}}}  
\newcommand{\Tw}[1]{\ensuremath{T_{w_{#1}}}} 
\newcommand{\Tg}{\ensuremath{T_{g}}\xspace}                    
\newcommand{\Memory}{\ensuremath{\mathcal{M}}} 
\newcommand{\chip}{\chi^{\prime}}     
\newcommand{\chipp}{\chi^{\prime\prime}} 
\newcommand{\freq}{\ensuremath{1\,\mathrm{Hz}} \xspace}
\usepackage{bm}
\usepackage{array}
\usepackage{xcolor}
\usepackage{float}
\usepackage [T1]{fontenc}
\usepackage [utf8]{inputenc}
\usepackage[colorlinks]{hyperref}

\usepackage[T1]{fontenc}
\usepackage[utf8]{inputenc}
\usepackage{graphicx}
\usepackage{amsmath}
\usepackage{amssymb}
\usepackage{xcolor}
\usepackage{xspace}
\usepackage{multirow}
\usepackage{hyperref}
\usepackage{dcolumn}
\makeatletter
\newcommand*{\balancecolsandclearpage}{%
  \close@column@grid
  \cleardoublepage
  \twocolumngrid
}
\makeatother
\graphicspath{{./figures/}{./}}

\AtBeginDocument{
                \newwrite\bibnotes
                \def\bibnotesext{prl_bib.bib}
                \immediate\openout\bibnotes=\jobname\bibnotesext
                \immediate\write\bibnotes{@CONTROL{REVTEX42Control}}
                \immediate\write\bibnotes{@CONTROL{apsrev42Control,author="08",editor="1",pages="1",title="0",year="1"}}
                \if@filesw
                \immediate\write\@auxout{\string\citation{apsrev42Control}}
                \fi
}

\begin{document}
\title{Memory and Rejuvenation in Glassy Systems\\}

\author{J. Freedberg} \email{freed114@umn.edu} \affiliation{School of Physics and Astronomy, The University of Minnesota, Minneapolis, Minnesota 55455, USA}

\author{W. Joe Meese}\affiliation{School of Physics and Astronomy, The University of Minnesota, Minneapolis, Minnesota 55455, USA}

\author{J. He}\affiliation{Department of Mechanical Engineering, The University of Texas at Austin, Austin, Texas 78712, USA}

\author{D.~ L.~ Schlagel} \affiliation{Division of Materials
  Science and Engineering, Ames National Laboratory, Ames, Iowa 50011, USA}

\author{E. Dan Dahlberg}\affiliation{School of Physics and Astronomy, The University of Minnesota, Minneapolis, Minnesota 55455, USA}

\author{R.~L.~Orbach}
\affiliation{Texas Materials Institute, The University of Texas at Austin,
  Austin, Texas  78712, USA}


\date{\today}
\begin{abstract}
The memory effect in a single crystal spin glass ($\mathrm{Cu}_{0.92}\mathrm{Mn}_{0.08}$) has been measured using \freq ac susceptibility techniques over a temperature range of $0.4 - 0.7 \, T_g$ and a model of the memory effect has been developed. A double-waiting-time protocol is carried out where the spin glass is first allowed to age at a temperature below $T_g$,  followed by a second aging at a lower temperature, \Tw{2}, after it has fully rejuvenated. The model is based on calculating typical coincident growth of correlated regions at the two temperatures.  It  accounts for the absolute magnitude of the memory effect as a function of both waiting times and temperatures.  The data can be explained by the  memory loss being  a function of the relative change in the correlated volume at the first waiting temperature because of the growth in the correlations at the second waiting temperature.
\end{abstract}

\maketitle

\section*{Introduction}The origin and nature of memory and rejuvenation in spin glasses has been the subject of experimental and theoretical investigations for over two decades 
\cite{jonason:98,vincent:07,dupuis:05,berthier:02, vincent:97,hammann:00, berthier:05b, vincent:95, refregier:87,vincent95b,jonason:00,sasaki:02,miyashita:01,yoshino:01,jonsson:04, vincent_spin_2018}.  When held at a temperature below the glass temperature $T_g$, the state of a spin glass is well known to age with time 
\cite{lundgren:83,chamberlin_1984,ocio_1985, vincent_spin_2018}.  Rejuvenation is the process where the spin glass appears to lose knowledge of its prior aging when the temperature is lowered. Memory, on the other hand, is displayed when the spin glass is reheated to the aging temperature and it recovers, at least partially, the aged state.  Although there is some agreement to the origin of the aging phenomena, rejuvenation and memory present a conundrum that has eluded a satisfactory simultaneous explanation
\cite{vincent_spin_2018}. The appearance of these effects together is central to understanding the spin glass state, in particular, how one can understand memory observed \textit{after} rejuvenation.  

This begs the question – if the spin glass appears to have ``forgotten'' it aged during rejuvenation, how can it then ``remember'' its previous cooling history?  Several explanations have been postulated, but never quantitatively tested experimentally \cite{jonason:98, lefloch:92, Jonnason:99, jonsson:04, vincent:00, jonason:00, miyashita:01}. As we will show, our double-waiting-time experiments and model answer this question.  

In this Letter, we quantify spin glass memory loss in a single crystal of $\mathrm{Cu}_{0.92}\mathrm{Mn}_{0.08}$ and present a simple physical model which accounts for our results (see Paga {\textit{et al.}} \cite{Paga:23} for a complementary numerical study of memory). It is expected that this picture may provide an explanation for memory in other glassy systems, including biopolymers, granular media, and structural glasses \cite{chen:23, kuersten:17,prados:14, tong:2023strain, jafari:2017,Saleh:2020}. 

\begin{figure}[h]
	\centering
	\includegraphics[width = \columnwidth]{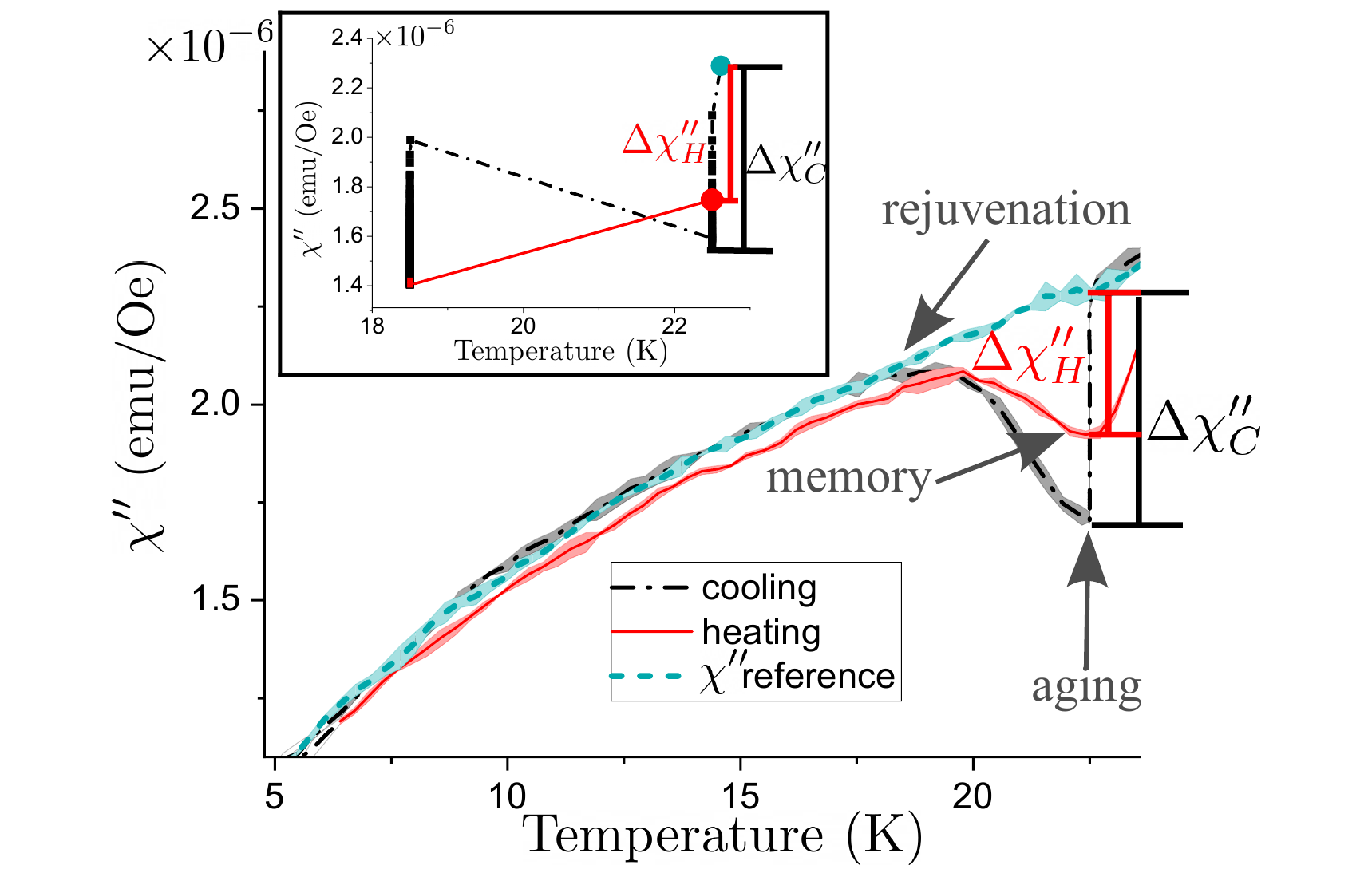}
	\caption{The imaginary part of the ac  magnetic susceptibility, $\chi^{\prime\prime}(\omega=2\pi\, \mathrm{Hz}, T)$.  The dashed data are the reference curve measured while continuously cooling the sample.  The dot-dash data include waiting at \Tw{1} $=22.5$ K for \tw{1} $= 1$ hour and  rejuvenation upon lowering the temperature. The solid curve is the heating data that exhibits the memory effect back at \Tw{1}.  All heating and cooling rates are 1 K/min. The inset shows a double-waiting-time experiment.  $\Delta \chipp_{C}$ is the change in susceptibility during aging and $\Delta\chipp_{H}$ is the difference from the reference curve upon heating.  For these data, the cooling and heating rate is 35 K/min between the two waiting temperatures.  } 
	\label{fig:ref_aging_rej_mem}
\end{figure}

Fig. \ref{fig:ref_aging_rej_mem} shows a canonical low frequency ac susceptibility measurement displaying these out-of-equilibrium phenomena \cite{jonason:98}. The reference curve, where no aging is exhibited, is plotted alongside the curves that exhibit aging, rejuvenation, and memory. As in previous work \cite{jonason:98, Jonnason:99}, we focus on the imaginary part of the ac susceptibility $\chipp(\omega)$,  because the size of these effects is more pronounced than in the real part,  $\chip(\omega)$  \footnote{Similar conclusions to the $\chipp(\omega)$ data can be drawn for  $\chip(\omega)$. }. Hereafter, we drop the explicit frequency dependence on the susceptibility $\chi = \chip + {\rm i}\chipp$.

A significant feature is  shown in Fig. \ref{fig:ref_aging_rej_mem}, where upon returning to $\Tw{1}$,  while the heating curve does have knowledge of the cooling curve, they do not lie on top of each other.  Importantly, the spin glass exhibits ``memory loss.'' Of the many previous works that see memory, some appear to see nearly perfect memory \cite{jonason:98, lefloch:92}, while others see memory loss  \cite{Jonnason:99, jonason:00, jonsson:04}. A common qualitative picture states that memory is an effect of spin glass droplets whose dynamics are organized in a hierarchy ordered by their size
\cite{jonason:98, jonason:00, Jonnason:99}. The distribution of droplet sizes corresponds to a distribution of different relaxation times measured in experiments. However, recent simulation results \cite{janus:21} indicate that temperature chaos drives rejuvenation as a random process of destroying locally correlated regions of spin \textit{overlap}, or the local Edwards-Anderson order parameter, rather than just affecting the spin configurations themselves. These differing explanations again lead us to question of how memory can \textit{follow} rejuvenation, given that the system has gone chaotic and therefore lost knowledge of its previously grown correlated state.

Due to recent progress in understanding rejuvenation as a consequence of temperature chaos \cite{janus:21}, we return to the observation of memory loss in spin glasses. We follow the so-called ``double memory'' experiments by Jonason \textit{{et al.}} and Lefloch \textit{{et al.}} \cite{jonason:00, lefloch:92}, and focus on tuning the memory effect between two aging intervals at different temperatures separated by rapid temperature changes. In a similar spirit, we dub our protocols ``double-waiting-time'' experiments.

We quantify the memory, \Memory, by computing the ratio
\begin{equation}
\Memory \equiv \frac{\Delta \chipp_H}{\Delta \chipp_C} = \frac{ \chipp_R(\Tw{1}) - \chipp(\tw{1} + \tw{2}, \Tw{1})}{ \chipp_R(\Tw{1}) - \chipp(\tw{1}, \Tw{1})}, \label{eq:Memory_Definition}
\end{equation}
shown pictorially in Fig. \ref{fig:ref_aging_rej_mem}. This quantity is a dimensionless parameter that compares the three measured susceptibilities.  The numerator, $\Delta \chipp_H$, is the difference between the reference curve $\chipp_R$ at \Tw{1} and the susceptibility upon returning to  \Tw{1} after aging for an additional \tw{2} at the lower temperature.  The denominator, $\Delta \chipp_C$, is the difference between the reference curve $\chi^{\prime\prime}_R$ at \Tw{1} and the dynamic susceptibility at \Tw{1} after aging for \tw{1}. Perfect memory ($\Memory = 1$) occurs  when $\chipp(\tw{1} + \tw{2}, \Tw{1}) = \chipp(\tw{1}, \Tw{1})$. If there is no memory ($\Memory = 0$), the heating curve follows the reference curve, $\chipp(\tw{1} + \tw{2}, \Tw{1}) = \chipp_R(\Tw{1})$.

We ascribe rejuvenation to temperature chaos, first described in Ref. \cite{bray:78}, and recently observed  \cite{Zhai:22} in a sample of $\mathrm{Cu}_{0.92}\mathrm{Mn}_{0.08}$ cut from the same boule as our sample.  As shown computationally in Refs. \cite{miyashita:01, janus:21}, when the temperature is lowered to \Tw{2} such that the spin glass has gone chaotic, then spin glass correlations develop over regions with a new length scale, with the growth of new correlations at $\Tw{2}$ occurring as if there were no correlations grown at \Tw{1}. We have ensured our temperature drop of $\Delta T \equiv \Tw{1} - \Tw{2} = 4\,\rm{K}$ is large enough to guarantee rejuvenation (see Fig. \ref{fig:ref_aging_rej_mem}).

From this argument, memory is an interplay between correlation lengths, as recently confirmed  \cite{baity_2023}.  We increase the correlation lengths by varying the waiting time. If \tw{1} (\tw{2}) increases while \tw{2} (\tw{1}) is fixed, more (less) memory is expected at \Tw{1}. This argument is akin to the ``temperature microscope model'' (TMM) of Bouchaud \textit{et al.} \cite{bouchaud:01} used to describe locally-ordered spin configurations within the droplet picture, as well as the phenomenological picture presented in Ref \cite{jonason:00}.  Importantly, the TMM implies that memory in the double-waiting-time experiments are a function of a single variable: the ratio between the dynamical correlation lengths grown at either temperature. Meanwhile, models like the ``ghost domain'' picture \cite{jonsson:04} are similar to the TMM, but have correlated growth with finite spatial extent.  Still other pictures, such as in Ref. \cite{bray:78} imply that \textbf{any} growth of \textbf{any} correlated regions at \Tw{2} will lead to memory loss. In what follows, we scrutinize these arguments by measuring how the memory, defined in Eq. \ref{eq:Memory_Definition}, varies throughout our double-waiting-time experiments.

\section*{Methods}The following describes our systematic, quantitative study of the memory effect. All ac susceptibility measurements were taken using a Magnetic Property Measuring System (MPMS) 3 \footnote{Quantum Design North America, 10307 Pacific Center Court San Diego, CA 92121}. The measurement frequency was 1 Hz, with a field amplitude of 10 Oe, sufficient to observe out-of-equilibrium effects while staying in the linear regime. The glass temperature was found through dc magnetization measurements to be $T_g = 41.6$ K \footnote{The value is determined by the ``onset of irreversibility'' \cite{vincent_spin_2018}}.

For the double-waiting-time experiments, we approximate a quench with a cooling rate of 35 K/min to reduce unintended aging effects. For consistency in the measurements, the temperature was allowed to settle at each waiting temperature before recording the first measurement. While the temperature change from \Tw{1} to \Tw{2} was only about 10 seconds, the first data point taken was around 100 seconds.  For each recorded point, 10 measurements were averaged.  First, the system was quenched  from  $T = 60\, \mathrm{K} > T_g$ to an aging temperature $\Tw{1} < \Tg$ and was allowed to relax for a time \tw{1}. Then, the system was quenched to a lower temperature  $\Tw{2} < \Tw{1}$  where it evolved for time \tw{2}. Finally, the system was rapidly heated back to \Tw{1} where the susceptibility was compared to the reference system (inset of Fig. \ref{fig:ref_aging_rej_mem}). 

 \begin{figure*}

	 \centering
	\includegraphics[width = \textwidth]{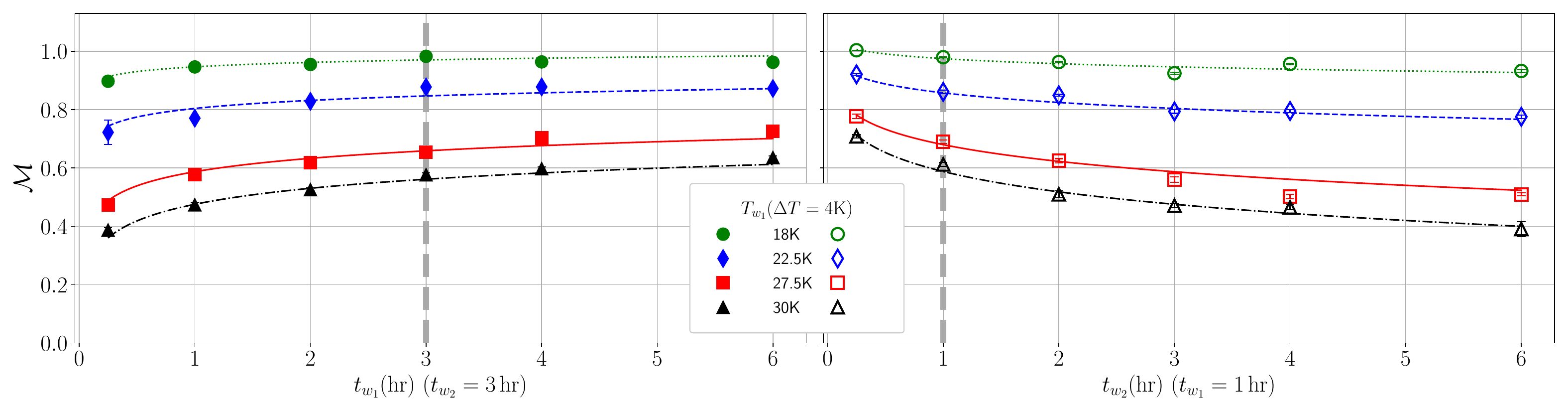}

	\caption{Memory versus waiting time with four different first waiting temperatures.  For  both (left) and (right), the first and second waiting temperatures are the same: \Tw{1} =  (18, 22.5, 27.5, and 30 K) and $\Tw{1} - \Tw{2} = 4\, {\rm K}$. In both cases either \tw{1} (left) or \tw{2} (right) were varied from $0 - 6$ hours (the horizontal axis), while the other waiting time was fixed at 3 and 1 hour, respectively.  Closed (open) markers with statistical error bars indicate a variation of \tw{1} ($\tw{2}$). The lines are predictions based our model described by Eq. \ref{eq:memory_final}. As discussed in the text, the vertical dashed lines are where we expect a crossing if memory only depends on the ratio of the spin glass correlation lengths.}
\label{fig:memory_vs_time_exp}
 \end{figure*}

\section*{Results}Our experiments were designed to test the effects of waiting time (and therefore the correlation length) on $\Memory$.  In Fig. \ref{fig:memory_vs_time_exp}, we see that the double-waiting-time protocol significantly impacts $\Memory$. The trend is clear -- the longer \tw{1}, the greater $\Memory$ is, but the longer \tw{2}, the smaller $\Memory$ is. Additionally, in experiments with a lower $\Tw{1}$, $\Memory$  is always larger than in ones that have a higher $\Tw{1}$. Furthermore, the fact that $\Memory$   \textit{increases} with \tw{1} means that the memory loss seen is more complicated than the picture presented in Ref. \cite{bray:78}, where any growth at \Tw{2} for a fully chaotic system is expected to decrease \Memory. Indeed, this increase suggests that the degree to which temperature chaos erases the spin glass' memory is a gradual, rather than an abrupt, process.  

There are at least two length scales at play – the size of correlations at \Tw{1}, and those at \Tw{2}. We estimate the correlation length based on the relationship developed by Kisker and Rieger 
\cite{kisker:96}
\begin{align}
\frac{\xi(t,T)}{a_0} = c_1 \left( \frac{t}{\tau_0}\right)^{c_2T/T_g}, \label{eq:corr_len}
\end{align}
where $a_0 = 6.6$\AA   \,\,is the average spacing between manganese ions, $\tau_0 \approx 2 \times 10^{-13} \,\mathrm{s}$ is the timescale of microscopic fluctuations, $c_1 \approx 1$  and $c_2 \approx 0.1$. These estimates have been compared to three experimentally extracted correlation lengths (starred points in Fig. \ref{fig:data_collapse}) using dc protocols pioneered by \cite{joh:99, zhai:19}.

At first glance, our experimental results in Fig. \ref{fig:memory_vs_time_exp}, are qualitatively consistent with the TMM, which posits that the memory loss is only controlled by the ratio $\alpha \equiv \xi_2/\xi_1$. An increase in $\xi_1$ ($\xi_2$) leads to a decrease (increase) in memory loss. However,  quantitatively, this is not borne out in the data. To illustrate this,  consider when $\tw{1} = \tw{2}$.  $\alpha$ will then depend only on the \textit{difference} between waiting temperatures, $\Delta T$, meaning the curves in Fig. \ref{fig:memory_vs_time_exp} should cross when $\tw{1}=\tw{2}$ since all the $\Delta T$s are identical (for more details, see the supplemental material in Ref. \cite{supplemental} ). Experimentally, however,  we find that \Memory \, varies by about $40\%$. Even when the temperature dependence of $c_2$ is taken into account, only a $2.4\%$ difference in \Memory \, is  expected.  This demonstrates that the physics of memory has a more intricate dependence on spin glass correlations than what is currently discussed in the literature. 

\section*{Discussion}To this end, we developed an experimentally-motivated model of our data whose derivation is given in the supplemental (Ref. \cite{supplemental}). We consider a growing correlated region of size $\xi_1$ at waiting temperature \Tw{1}, encapsulated by a volume $V_{\ell}=\ell \ell^2_\perp$. Here, we take $\ell$  and $\ell_\perp$ as length scales that are parallel and perpendicular to the ac field, and require that they are large enough to consider different volumes of size $V_{\ell}$ as statistically independent. In equilibrium, and in the absence of a real-space anisotropy, we would expect that these scales are both equal to a static, isotropic correlation length which we assume is always larger than the dynamical correlation length in our experiments. 

Next, we consider a quench to \Tw{2}. Because we have ensured full rejuvenation, the spin glass energy landscape at \Tw{2} does not have the same set of minima as it does at \Tw{1}. We model this as a single correlated region of volume $V_{\xi_1}$ growing within $V_{\ell}$, with secondary correlated regions of volume $V_{\xi_2}$ growing independently from the first. Thus, upon quenching to \Tw{2}, the new regions can appear anywhere within $V_\ell$ with an assumed uniform probability. If any new growth at \Tw{2} coincides with the original correlated region during the waiting time \tw{2}, we assume memory is reduced upon returning to \Tw{1}. 

Within this model, memory loss is the average relative change in $V_{\xi_1}$ within $V_{\ell}$ after randomly developing new correlated regions at \Tw{2}. Thus, we must compute $\overline{\Delta V}/V_{\xi_1}$, where the overbar denotes statistical averaging over all independent volumes of size $V_\ell$. Further details can be found in the supplemental \cite{supplemental} and Ref. \cite{sphere_sphere} therein, but the main point is that the average change in volume $\overline{\Delta V}$ depends on where $V_{\xi_2}$ grows: within, at the edge of, or outside $V_{\xi_1}$.  The final expression for a spherical correlated volume, $\overline{\Delta V}/V_{\xi_1}$, is proportional to  
\begin{align}
    \frac{\overline{\Delta V}}{V_{\xi_1}} \propto \frac{V_{\xi_1}}{V_\ell} \left( \frac{\xi_2}{\xi_1} \right)^3 = \frac{V_{\xi_1}}{V_\ell}\alpha^3 . \label{eq:generic_3D_DeltaV_over_V}
\end{align}
Comparing Eq.\ref{eq:3D_DeltaV_over_V} to Fig. \ref{fig:memory_vs_time_exp}, we identify two cases that help ascertain the behaviors of $\ell$ and $\ell_\perp$. The first case would be if there is no independent length scale at the first waiting temperature other than $\xi_1$, implying both $\ell$  and $\ell_\perp  \sim \xi_1$. Thus, Eq. \ref{eq:generic_3D_DeltaV_over_V} predicts that memory loss only depends on $\alpha$ (TMM). This is  incompatible with the data shown in Fig. \ref{fig:memory_vs_time_exp}, as discussed above. In the second case, there is still real-space isotropy, $\ell = \ell_\perp$, but both $\ell_{\perp}$ and  $\ell$ are independent of $\xi_1$. In this case, Eq. \ref{eq:generic_3D_DeltaV_over_V} implies that memory loss does not depend on $\xi_1$ at all, again incompatible with the experimental results -- were this true, each curve in Fig. \ref{fig:memory_vs_time_exp} (left) would be constant. Hence, from our experimental data, we conclude that only one length scale must indeed scale with $\xi_1$, while the other is independent of it. 

By substitution into Eq.  \ref{eq:generic_3D_DeltaV_over_V}, we have
\begin{align}
    \frac{\overline{\Delta V}}{V_{\xi_1}} \propto \frac{4\pi}{3}\frac{\xi_1}{\ell} \left( \frac{\xi_2}{\xi_1} \right)^3 = \frac{4\pi}{3}\frac{\xi_1}{\ell}\alpha^3. \label{eq:3D_DeltaV_over_V}
\end{align}

\begin{figure}
    \centering
    \includegraphics[width = \columnwidth]{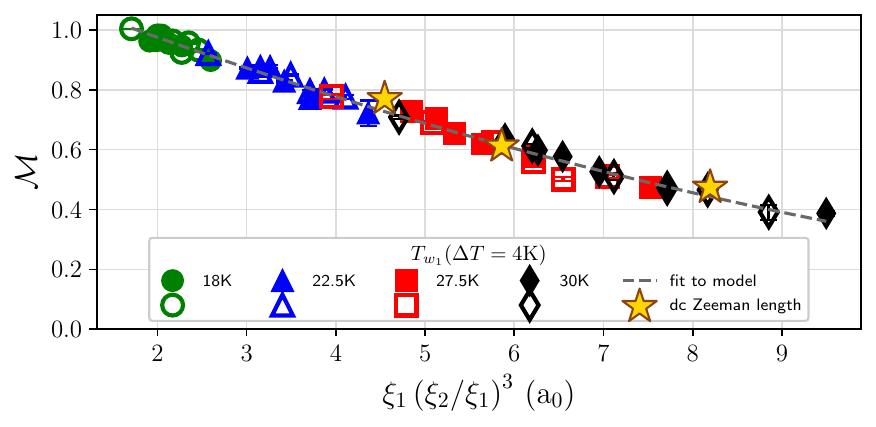}
    \caption{The data from Fig. \ref{fig:memory_vs_time_exp} \textit{collapses} as a function of memory loss modeled in Eq. \ref{eq:3D_DeltaV_over_V}.  Additionally, the three starred data points use correlation lengths extracted from dc experiments with the ac values of memory. \Tw{1} (18, 22.5, 27.5, and 30 K) and $\Tw{1} - \Tw{2} = 4\, {\rm K}$. In both cases either the first (closed markers) or second (open markers) waiting times are varied from $0 - 6$ hours, while the other waiting time is fixed at 3 (1) hours. }
    \label{fig:data_collapse}
\end{figure}

With this expression, we combine all data from Fig.  \ref{fig:memory_vs_time_exp}  into a single plot shown in Fig. \ref{fig:data_collapse}; with the measured memory being the vertical axis, and the calculated value of $\xi_1\alpha^3$  being the horizontal. We see clear evidence of data collapse from all of the double-waiting-time experiments, indicating that the memory effect is not solely a scaling function of $\alpha$, but rather memory depends on \textit{both} length scales explicitly. The only way the dimensionless quantity of memory can be a function of both lengths is if there is at least another independent length scale present: $\ell$. Our model provides an interpretation for this length. Its size, compared to $\xi_1$, controls how probable it is for new correlated regions grown at \Tw{2} to coincide with those at \Tw{1}. For it is only growth at \Tw{2} that coincides with the original growth at \Tw{1}, rather than any growth at all, that leads to memory loss.

We emphasize that the collapse shown in Fig. \ref{fig:data_collapse} comes from 52 independent double-waiting-time experiments \footnote{48 data points and four additional averages}, and the horizontal axis represents a protocol-dependent, \textit{calculated} value from Eq. \ref{eq:3D_DeltaV_over_V}. The fact there is a distinct data collapse represents agreement between these independent trials and implies the existence of another physical length scale relevant for memory. This is the additional structure that has been missing in our understanding of the memory effect.

Now, we revisit the conundrum where some spin glasses exhibit significant memory loss (typically metallic spin glasses) while others do not (typically insulating spin glasses). If a system has $\xi_1/\ell \rightarrow 0 $, we expect nearly perfect memory regardless of correlated growth at \Tw{2}. Meanwhile, whenever $\xi_1/\ell$ is sizeable, there will always be a substantial amount of memory loss. In both cases, random competition between the independent growth of $\xi_1$ and $\xi_2$ drives memory loss, and $\ell$ controls the severity. 

We can be more precise with our modeling. As discussed in Refs.  \cite{zhai:17, kenning:18, janus:18, zhai-janus:20a}, an increase in $\xi$ is attributed to growing spin glass order. In the case of ac susceptibility measurements, this is seen in the reduction of $\chipp$.  As derived in the supplemental \cite{supplemental}, the relation between $\mathcal{M}$ and  $\overline{\Delta V}/V_{\xi_1}$ can then be shown to have the form
\begin{align}
    \mathcal{M} = w \left[ 1 - \frac{4\pi}{3} \,\frac{\xi_1}{\ell} \,\left(\frac{\xi_2}{\xi_1}\right)^3 \right]^{p/d}, \label{eq:memory_final}
\end{align}
where $w \sim \mathrm{O}(1)$, $\ell$, and $p$  are unknown fitting parameters. For Fig. \ref{fig:data_collapse}, the values of  $w$, $\ell$, and  $p$ are found to be $w = 1.2 \pm 0.02, \, \ell = 100 \pm 24\, a_0$, and $p/d = 2.4 \pm 0.7$, where $d = 3$ is the spatial dimension. 

This predicted fit is shown in Fig.  \ref{fig:data_collapse}, and we find that the data in Fig. \ref{fig:memory_vs_time_exp} are  well-represented by this model. Notably, the same three fitting parameters are used for all 52 independent trials of the eight curves in Fig.  \ref{fig:memory_vs_time_exp}.

\section*{Conclusions}  Using quantitative measures of memory loss in spin glass systems, we have elucidated the mechanism responsible for the memory effect. By performing a ``double-waiting-time'' experiment, and varying either \tw{1} or \tw{2} while holding the other fixed, we have shown that we can tune the amount of memory loss. This has allowed us to quantitatively test previously proposed qualitative explanations of memory loss and develop an experimental model which explains our results. By modeling the amount of memory retained as a function of dynamical correlation length growth, we find that an additional spatial length scale, $\ell$ controls the impact that independent, but coincident, growth of correlations at \Tw{2} have on established correlated regions grown at \Tw{1}.

Our modeling relies on the existence of separate, uncorrelated anisotropic volumes of size $V_\ell=\ell \ell_\perp^2$, where $\ell_\perp$, measured in the plane normal to the ac field, scales with the dynamical correlation length grown at the first waiting temperature $\xi_1(\tw{1},\Tw{1})$, and  $\ell$ is independent of it. While the cause of the spatial anisotropy is unclear, our data (Fig. \ref{fig:memory_vs_time_exp}), and subsequent collapse of all 52 independent trials (Fig. \ref{fig:data_collapse}), strongly support its presence. There are several mechanisms which could generate a spin-space anisotropy in spin glasses like an external magnetic field acting as an effective uniaxial random field \cite{Aharony_1978}, or magnetic anisotropies inducing chiral order \cite{Kawamura_1992}. At present, it is unclear how these would present in real-space and will be the subject of further study, but such spin-space mechanisms exist.  For example, extended defects are known to play a similar role in disordered magnetic systems \cite{Cardy_1982, Prudnikov_1983, Prudnikov_1984, Holovatch_2005, Holovatch_2015}. Our model assumes the spin-space anisotropy produced by the ac field generates an equivalent real-space anisotropy.\\
\indent We emphasize that the spin glass correlated volumes as described in this Letter are not spin-spin correlations as suggested in the droplet picture \cite{fisher:88}. Instead, given the recent numerical results on temperature-chaos-driven rejuvenation \cite{janus:21}, these correlated volumes are grown in the local Edwards-Anderson overlap. Crucially, this allows a description of memory \textit{following} rejuvenation. When the temperature is lowered, it is the Edwards-Anderson overlap, rather than the spin configurations which are frozen in, or ``imprinted.'' Once the system goes chaotic, the growth of correlations at either waiting temperature will be independent of the growth of correlations at the other waiting temperature. However, the correlations at the lower temperature could form in the same location in real-space as the original correlated regions. When this coincident growth occurs, the volume of the original correlated region decreases, leading to the memory loss observed.  We anticipate this study will allow for quantitative comparisons of the memory effect in other glassy systems \cite{bellon_2000}.

\begin{acknowledgments}
The authors wish to thank the Janus II collaboration, and especially Ilaria Paga, Victor Martin-Mayor, Juan Ruiz-Lorenzo, and Enzo Marinari for their collaboration. Additionally, we thank Virginia Gali and Aset Khakimzhan for their helpful review of this manuscript, as well as Kan-Ting Tsai,  Gregory G.  Kenning, and Rafael M. Fernandes for helpful discussions. The experimental  work was supported by the U.S. Department of Energy, Office of Basic Energy Sciences, Division of Materials Science and Engineering, under Award No. DE-SC0013599. WJM was supported by the U. S. Department of Energy, Office of Science, Basic Energy Sciences, Materials Sciences and Engineering Division, under Award No. DE-SC0020045.  Crystal growth of $\mathrm{Cu}_{0.92}\mathrm{Mn}_{0.08}$ sample was performed by Deborah L. Schlagel at the Materials Preparation Center, Ames National Laboratory, USDOE and supported by the Department of Energy-Basic Energy Sciences under Contract No. DE-AC02-07CH11358. All experiments were carried out at the Institute for Rock Magnetism (IRM) at the University of Minnesota which is supported by the National Science Foundation, Division of Earth Sciences, Instrumentation and Facilities under Award No. 2153786. The authors wish to thank Peat Solheid, Maxwell Brown, Dario Bilardello, Joshua Feinberg, and Bruce Moskowitz from the IRM for much equipment time, training, and access to this instrument.
\end{acknowledgments}

\bibliography{prl_bib.bib} 
\end{document}


\title{Supplementary Information. On the Nature of Memory and Rejuvenation in Glassy Systems}

\author{J. Freedberg} \email{freed114@umn.edu} \affiliation{School of Physics and Astronomy, The University of Minnesota, Minneapolis, Minnesota 55455, USA}

\author{W. Joe Meese}\affiliation{School of Physics and Astronomy, The University of Minnesota, Minneapolis, Minnesota 55455, USA}

\author{J. He}\affiliation{Department of Mechanical Engineering, The University of Texas at Austin, Austin, Texas 78712, USA}

\author{D.~ L.~ Schlagel} \affiliation{Division of Materials
  Science and Engineering, Ames Laboratory, Ames, Iowa 50011, USA}

\author{E. Dan Dahlberg}\affiliation{School of Physics and Astronomy, The University of Minnesota, Minneapolis, Minnesota 55455, USA}

\author{R.~L.~Orbach}
\affiliation{Texas Materials Institute, The University of Texas at Austin,
  Austin, Texas  78712, USA}

\date{\today}   

\maketitle

\onecolumngrid\

\maketitle

\section{Numerical estimation of the spin glass correlation length \label{est_xi}}

To estimate the correlation length, we use the Kisker-Rieger growth law \cite{kisker:96}
\begin{align}
\xi(t_w, T_w) = c_1 a_0\left( \frac{t_w}{\tau_0}\right)^{c_2 T_w/T_g},  \label{eq:corr_len}  
\end{align}
where $a_0 = 6.6$ \AA\space is the microscopic distance between Mn ions,  $c_1 = 1$, $ c_2$ is determined by simulations to be  $\sim 0.1$  \cite{kisker:96},  $T_g = 41.6$ K is the glass temperature, and $\tau_0 = \hbar/k_B T_g\approx 1.86 \times 10^{-13} {\rm s}$  is the timescale of  microscopic fluctuations as determined by the energy-time uncertainty principle. We note the value of $c_2$ has been studied extensively both experimentally and computationally \cite{kisker:96, janus:18,zhai:17,zhai:19, kenning:18,joh:99}, and while these values are all about $0.1$, there are differences between them which are temperature \textit{and} sample dependent \footnote{The sample-dependence of $c_2$ is seen between polycrystalline and single-crystal samples of the same Mn concentration, even when the correlation length is small compared to the crystallite size in the former case \cite{joh:99, zhai:19}.}. For the purposes of this letter, we compare our approximated values for $\xi$ to three experimentally extracted lengths from dc experiments which we use as a gauge to determine $\xi_1\alpha^3$ for 3 double-waiting-time experiments. This allowed us to determine the best choice of $c_2$ from the available values in the literature, tuned specifically for the time and temperatures in our experiments. The closest agreement comes from the simulations conducted in Ref. \cite{kisker:96}, using a $c_2=c_2(T)$ of  $0.117 - 0.104$ for a temperature range $8.3 - 29.1$ K, respectively. Values of $c_2$ within this temperature interval are linearly extrapolated. 

\section{Evidence of a third length scale}
The memory effect, as previously described as a function of only two correlation lengths, cannot capture the memory loss  reported in this Letter. Indeed, this can be understood by directly writing the ratio of the Kisker-Reiger correlation lengths (from Eq. \ref{eq:corr_len})
$$\alpha \equiv\frac{\xi_2}{\xi_1} = \frac{c_1a_0\left(t_{w2}/\tau_0\right)^{c_2 T_2/T_g}}{c_1a_0\left(t_{w1}/\tau_0\right)^{c_2 T_1/T_g}}$$  In our experiments, $T_2 = T_1 - \Delta T$, with $\Delta T = 4\,{\rm K}$. To demonstrate the problem with  memory loss being solely a function of $\alpha$, we first assume that  $c_1, a_0,$ and $c_2$ are constants in temperature. As stated in \ref{est_xi}, it is well-known in the literature that the numerical value of $c_2$ is temperature-dependent, though its absolute value ranges on the (maximal) interval of 0.091 to 0.124 over the temperature range we measured \cite{janus:18}\footnote{The simulated value of $c_2$ also depends on temperature in \cite{kisker:96}, however this range is smaller than the interval we used in this estimate.}. We revisit the case of a changing $c_2$  after the case where we assume it is a constant, mean value of $0.11$. We recover the expression
\begin{align}\alpha = \left(t_{w2}/\tau_0\right)^{-c_2\Delta T /T_g}\left(t_{w2}/t_{w1}\right)^{c_2 T_1/T_g}. \label{eq:alpha_not_eq_time}\end{align}
From Eq. \ref{eq:alpha_not_eq_time}, there is one special case which demonstrates how this picture breaks down with our experimental data: $t_{w1} = t_{w2}$. In this case, the ratio becomes

\begin{align}\alpha\left(t_{w1}=t_{w2}\right) \equiv \alpha_{\star}= \left(t_{w2}/\tau_0\right)^{-c_2\Delta T /T_g}.\label{eq:same_tw}\end{align}
Clearly, in this special case, there is no temperature dependence in $\alpha_\star$, and therefore we would expect memory to be a constant value for each pair of waiting temperatures we measured, if memory indeed only depends on the ratio of the two correlation lengths. In other words, the memory $\mathcal{M}$ in Fig. 2 of the main text should collapse to a single point along the vertical dashed line in both subplots \footnote{As will be shown in Fig. \ref{fig:fixed_alphas} of the supplemental, the same substantial memory loss is typical over multiple trials when the ratio of correlation lengths $\alpha$ is held fixed.}.  But, this is not what is observed. In the figure, it is evident that $\mathcal{M}$ ranges from almost $1$ to slightly less than $0.6$, corresponding to a 40\% change in memory, well outside the experimental error bars.

This quick calculation was founded on $c_2$ being temperature-independent, which it is not. However, because it does not vary in magnitude substantially over the temperature range we considered, we can improve our estimate of the expected memory loss from the literature by considering the effect of small variations on the coefficient $\alpha_{\star}$.  This is valid because the exponent $c_2 T/T_g$ is always small ($\sim 0.01-0.1$). We write
\begin{align}
    \frac{\left\vert\delta \alpha_\star\right\vert}{\alpha_\star} &= \left|\frac{1}{\alpha_\star}\frac{\partial \alpha\star}{\partial c_2}\delta c_2\right| = \left|\frac{\partial \ln( \alpha\star)}{\partial c_2}\delta c_2\right|,
\end{align}
as the linearized relative change of $\alpha_\star$.  By inserting Eq. \ref{eq:same_tw} and taking the appropriate derivatives, we find
\begin{align}
    \frac{\left\vert\delta \alpha_\star\right\vert}{\alpha_\star} = \frac{\Delta T}{T_g}\ln\left(\frac{t_{w2}}{\tau_0}\right)\delta c_2, \label{delta_alpha_diff}
\end{align}
where we note that the term $\Delta T/T_g = 0.096$ and $\ln\left(\frac{t_{w2}}{\tau_0}\right) \approx 38.6$ for the set of experiments when $t_{w1}=t_{w2}= 3\,\rm{hours}$. Considering the absolute change in $c_2$ over our temperature range is  $\delta c_2 = 0.0165$  \cite{janus:18}, we then estimate the relative change in $\alpha_\star$  to be 
\begin{align}
    \frac{\left\vert\delta \alpha_\star\right\vert}{\alpha_\star} \approx 6.13\%,
\end{align}
which is still unable to account for the nearly $40\%$ difference observed experimentally.  Based on this calculation, memory loss in spin glass depends on more than just $\alpha$. We note that this  calculation constitutes an overestimate of the relative change, as we deliberately chose the largest range of temperature-dependent values of $c_2$ from the literature, and focused on the larger value of $t_{w2}$  (Fig. 2a of the main text), to exaggerate any change in $\alpha_\star$. If we had, for example, used the $c_2$ range taken from \cite{kisker:96}, which we found in \ref{est_xi} better represents our data, then the relative change  $\vert\delta\alpha_\star\vert/\alpha_\star \approx 2.41\%$. We conclude there is more to the memory effect observed in our experiments. 

In the section that follows, we derive our experimental model described in the main text which is able to account for our experimental results.

\section{ Modeling Memory Loss}

Here, we provide a derivation of the model describing memory loss presented in the main text. We start with  how we compare our ac susceptibility results to the dynamical correlation length $\xi(t_w, T_w)$. When ac susceptibility measurements are taken, $\chi$  \textit{decreases} while the temperature is held constant ($T_w$) for a waiting time $t_w$. Previous research has shown \cite{zhai-janus:20a} that during this time, the dynamical correlation length \textit{grows}, and thus, we assume a power-law relationship between the susceptibility decay, $\Delta \chi(t)$, and the dynamical correlation length. Written explicitly, we take $\Delta\chi(t, T) \sim \xi^p(t, T)$, where $p>0$.

Our main result is shown later (Eq. \ref{eq:final_anisotropic_result}) and finds that the ``memory loss'' exhibited by the spin glass during a double-waiting-time experiment is a function of the ratio of two important volumes. The first volume of course is  that subtended by the correlation length grown during the first aging step $V_{\xi_1}$. The second is the average \textit{change} in this volume $\overline{\Delta V}$ due to the independent growth of new correlated regions of radius $\xi_2$ during the second aging step after the spin glass has fully rejuvenated.  From this, we can derive an expression for the  memory, $\mathcal{M}$. We can relate the correlation length to the correlated volume by $\xi^p(t) \propto V^{p/d}(t)$, where $d = 3$ is the spatial dimension. 

In the main text, we defined memory $\mathcal{M}$ experimentally as $\Delta \chi_H/\Delta \chi_C$, where the $H$ and $C$ subscripts refer to measurements taken while heating or cooling, respectively.  By relating the susceptibility decays to their corresponding correlated volumes, we find that memory scales as 

\begin{align}
    \mathcal{M} &\equiv \frac{\Delta\chi_H}{\Delta \chi_C} \nonumber
    \\
    &\sim \frac{\left(V_{\xi_1} - \overline{\Delta V}\right)^{p/d}}{V_{\xi_1}^{p/d}} \nonumber\\
    & = \left(1-\frac{\overline{\Delta V}}{V_{\xi_1}} \right)^{p/d} .
\end{align}
These arguments predict that the spin glass memory $\mathcal{M}$ has a functional dependence on the memory loss $\overline{\Delta V}/V_{\xi_1}$ as 
\begin{align}
    \mathcal{M} = w \left(1-\frac{\overline{\Delta V}}{V_{\xi_1}} \right)^{p/d},\label{eq:M_expression}
\end{align}
where $w\sim {\rm O}(1)$ and $p$  are unknown fitting parameters. 

At this point, we focus on estimating the memory loss in the double-waiting-time experiments, which we ultimately arrive at as our Eq. \ref{eq:final_anisotropic_result}. We assume that whenever there is coincident growth between correlated volumes at the two different temperatures, hereby referred to as ``overlap,'' then the new correlated growth contributes to memory loss. In the following analysis, we simplify the geometry of the spin-glass correlated regions  by assuming a spherical structure in $d = 3$ dimensions.

To begin, we first consider a single cluster of radius $\xi_1 \equiv \xi(t_{w_1}, T_{w_1})$ centered within a volume subtended length scales $\ell_\parallel$  and $\ell_\perp$,  in the longitudinal and transverse directions to the ac field, respectively. We assume that these scales are large enough to separate the spin glass sample into statistically independent regions, requiring that $\ell_\parallel$ and $\ell_\perp$ are larger than $\xi_1$. Note that since $\xi_1$ is time-dependent, then both of these length scales may develop a time-dependence as well. We assume that either length scale $\ell_a$  \textit{dynamically scales} as $\xi_1$  if $\ell_a/ \xi_1 = {\rm const}$ for all time, with $a\in \{\parallel, \perp\}$.  For the time being, we remain agnostic about the dynamical dependences, however. Thus, we consider the growth of only a single correlated volume at the first waiting temperature within a subtending volume $V_{\ell} = \ell_\parallel\ell_\perp^2$.  

\begin{figure}
    \centering
    \includegraphics[width = 0.5\columnwidth]{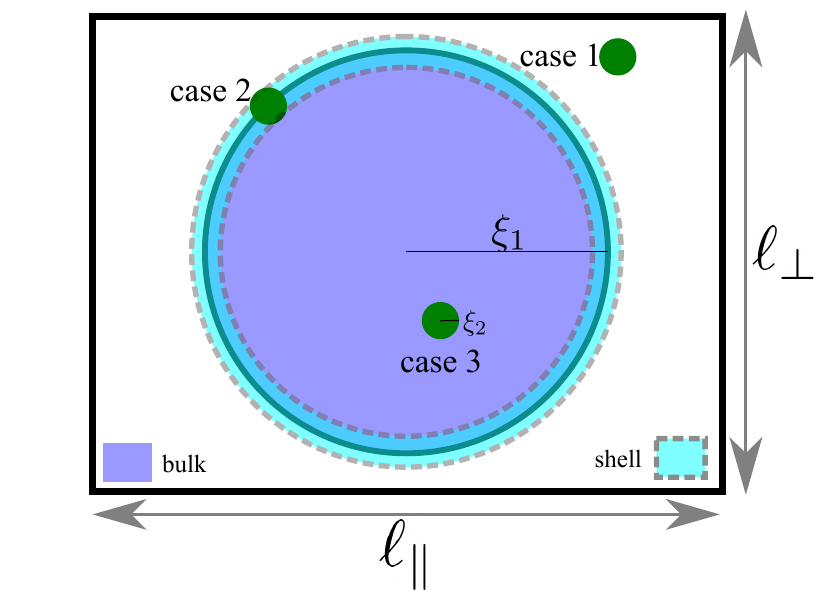}
    \caption{Assuming $\xi_1$ is encapsulated in a volume with sides of length $\ell_{\parallel}$ and $\ell_{\perp}$, a cluster of size $\xi_2$ can grow in these three regions with uniform probability. }
    \label{fig:model_cases}
\end{figure}

We next consider a second cluster of radius $\xi_2 \equiv \xi(t_{w_2}, T_{w_2}) < \xi_1$ which develops after the spin glass fully rejuvenates. In this case, the spin glass begins to develop correlations independently of its history, and therefore we assume that the new correlated regions grow randomly within the volume $V_{\ell}$. For example, we may take the new correlated regions grown as a sphere of size volume $\xi_2^3$  centered at position $\vb*{r}_0=(x_0, y_0, z_0)$. The average volume of the overlap between the new ($NC$) and  original  ($OC$) correlated volumes is $\overline{\Delta V}$.  These sets are defined by

\begin{align}
    OC &= \left\{ (x, y, z) \in V_\ell : \; x^2 + y^2 + z^2 \leq \xi_1^2 \right\},
    \\
    NC &= \left\{ (x, y, z) \in V_\ell : (x - x_0)^2 + (y - y_0)^2 + (z - z_0)^2 \leq \xi_2^2 \right\}.
\end{align}
Since the new cluster is defined by its center $r_0 = (x_0,y_0, z_0)$ and radius  $\vb*{r}$, we see that are three possible cases to consider as shown in Fig. \ref{fig:model_cases}. They follow as
\begin{enumerate}
    \item $NC$ is outside of $OC$. $\left\{ (x_0, y_0, z_0): \; x_0^2 + y_0^2 + z_0^2 > (\xi_1 + \xi_2)^2 \right\}$ \label{case:3D_NC_outside_eq}
    \item $NC$ is on the shell of $OC$. $\left\{ (x_0, y_0, z_0): \; (\xi_1 - \xi_2)^2 \leq x_0^2 + y_0^2 + z_0^2 \leq (\xi_1 + \xi_2)^2 \right\}$ \label{case:3D_NC_shell_eq}
    \item $NC$ is within the bulk of $OC$. $\left\{ (x_0, y_0, z_0): \; x_0^2 + y_0^2 + z_0^2 < (\xi_1 - \xi_2)^2 \right\}$ \label{case:3D_NC_bulk_eq}
\end{enumerate}
In case \ref{case:3D_NC_outside_eq}, the overlap is zero, and this term contributes nothing to $\overline{\Delta V}$.  In case \ref{case:3D_NC_bulk_eq}, the overlap is a constant and equal to the volume of $NC$: $V_{\xi_2} = 4\pi\xi_2^3/3$. In the intermediate case, case \ref{case:3D_NC_shell_eq}, the overlap changes as a function of position. These cases are illustrated in the cartoon in Fig. \ref{fig:model_cases}.\\

Thus, since the contribution to the overlap from case \ref{case:3D_NC_outside_eq}  is $0$, we simply need to look at cases \ref{case:3D_NC_shell_eq} and \ref{case:3D_NC_bulk_eq}. Therefore, the expected overlap $\overline{\Delta V}$ is 
\begin{align}
    \overline{\Delta V} = p_3 V_{\xi_2} + h_{3D}.
\end{align}
The probability $p_3$ is by the ratio of the volume of the bulk of $OC$ to that of the encapsulating volume $V_{\ell}$, represented by  case \ref{case:3D_NC_bulk_eq} in Fig. \ref{fig:model_cases}:
\begin{align}
    p_3 = \frac{\frac{4\pi}{3}\, (\xi_1 - \xi_2)^3}{V_\ell} = \frac{V_{\xi_1}(1-\alpha)^3}{V_\ell},
\end{align}
where the ratio between the correlation lengths $\alpha \equiv \xi_2/ \xi_1$.

Meanwhile, $h_{3D}$ follows as the expected overlap of the two correlated regions when $NC$ is randomly placed on the shell of $OC$ with uniform probability as pictured in case \ref{case:3D_NC_shell_eq}  in Fig. \ref{fig:model_cases}. We therefore are left to compute 
\begin{align}
    h_{3D} &= \sum_{(x_0, y_0, z_0) \in {\rm Shell}(OC)} p(x_0, y_0, z_0) \eta_{3D}(x_0, y_0, z_0) \nonumber
    \\
    &= \frac{4\pi}{V_\ell} \int_{\xi_1-\xi_2}^{\xi_1+\xi_2} {\rm d}r_0\, r_0^2 \eta(r_0),
\end{align}
where $\eta(r_0)$ is the overlap of two spheres \cite{sphere_sphere} given by 
\begin{align}
    \eta(r_0) &=  \frac{\pi}{12 r_0} \left( \xi_1 + \xi_2 - r_0 \right)^2 \left( r_0^2 + 2r_0\xi_2 -3\xi_2^2 + 2r_0\xi_1 + 6\xi_2\xi_1 -3\xi_1^2 \right),
    \end{align}
 and as a function of the dimensionless parameter $u \equiv r_0/\xi_1$:
    \begin{align}
    \eta(u) &=  \frac{\pi \xi_1^3}{12u} \left(1 + \alpha -u\right)^2 \left[ u^2 + 2(1+\alpha)u - 3(1-\alpha)^2 \right].
\end{align}
This integration yields an analytic form as a function of $\alpha$ given by 
\begin{align}
    h_{3D} = \frac{16\pi^2 \xi_1^6}{3V_\ell} \alpha^4 \left( 1 - \alpha + \frac{1}{3}\alpha^2 \right) = 4\pi V_{\xi_1} \left( \frac{\xi_1}{\ell} \right)^3 \alpha^4 \left( 1 - \alpha + \frac{1}{3}\alpha^2 \right).
\end{align}
By combining these results, we have that 
\begin{align}
    \overline{\Delta V}  &= \frac{4\pi}{3}\frac{V_{\xi_1}}{V_\ell}\left[ \xi_2^3(1-\alpha)^3 + \xi_1^3\alpha^4\left(3-3\alpha + \alpha^2 \right)\right]
\end{align}
From these expressions, we have that the ratio of the expected change in correlated volume to the original correlated volume is 
\begin{align}
    \frac{\overline{\Delta V}}{V_{\xi_1}} &= \frac{4\pi}{3}\frac{\xi_1^3}{V_\ell}\alpha^3. \label{eq:final_isotropic_result}
\end{align}

The Eq. above is consistent with several different functional behaviors for memory loss, depending on the size and dynamics of $V_\ell$. For example, if we were to take $V_\ell \propto \xi_1^3$, implying that both lengths $\ell_\parallel$ and $\ell_\perp$ dynamically scale as $\xi_1$, then it follows that the memory loss $\overline{\Delta V}/V_{\xi_1}$ would be simply proportional to $\alpha^3 = (\xi_2/\xi_1)^3$. This would be the first relevant case, and it is the one discussed by Bouchaud \textit{et al.}  and Lefloch \textit{et al.} \cite{bouchaud:01, lefloch:92}. Another relevant case would be if \emph{both} length scales $\ell_\parallel$ and $\ell_\perp$ did not dynamically scale with $\xi_1$, but they were isotropic in the sense that $\ell_\parallel=\ell_\perp$. If this were the case, then Eq. \ref{eq:final_isotropic_result} would predict that memory loss only would depend on $\xi_2$, as the $\xi_1$  dependence would cancel. To ascertain the dynamical scaling of $\ell_\parallel$ and $\ell_\perp$ relative to $\xi_1$, we directly compare both cases with our experimental data. In what follows, we argue that both of these scenarios are incompatible with our data.

\begin{figure}
    \centering
    \includegraphics[width=
0.5\columnwidth]{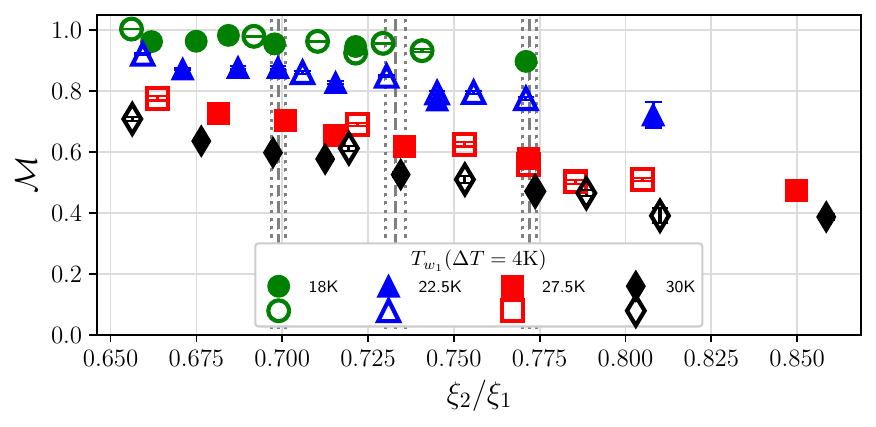}
    \caption{$\mathcal{M}$ plotted as a function of $\alpha \equiv \xi_2/\xi_1$ for the data shown in Fig. 2 of the main text.  Clearly, the memory is dependent on something other than $\alpha$.  Closed (open) markers come from trials where the first (second) waiting time is varied. Statistical error bars are present on each data point, though they are typically smaller than the marker.}
    \label{fig:M_vs_alpha}
\end{figure}

We first compare the former case -- $V_\ell \propto \xi_1^3$ -- to our experimental data. To do so, we compute the values of the Kisker-Rieger dynamical correlation length \cite{kisker:96} at both waiting temperatures and waiting times for each of our individual double-waiting-time experiments shown in Fig. 2 of the main text. This allows us then to compute $\alpha = \xi_2/\xi_1$  for each experiment. We then show the memory $\Memory$ as a function of each of these $\alpha$  values in a scatter plot in Fig. \ref{fig:M_vs_alpha}. We notice that, while there is a trend between trials with the same first waiting temperature, \Tw{1}, there is still a wide scatter present between all trials with different \Tw{1}.  To make this point explicit, let us consider several values of constant $\alpha$, which are marked by the dashed vertical lines in Fig.  \ref{fig:M_vs_alpha} \footnote{To encapsulate several data points, we actually consider a small range of $\alpha$ values, the width of which is denoted by the adjacent dotted lines.}.  The spread in the values of $\Memory$ for identical values of $\alpha$ indicates memory loss cannot be solely a function of $\alpha$.  This means $V_{\ell}$ cannot be proportional to $\xi_1^3$ and therefore there must be another length scale at \Tw{1} relevant for memory.

\begin{figure}
    \centering
    \includegraphics[width=
0.5\columnwidth]{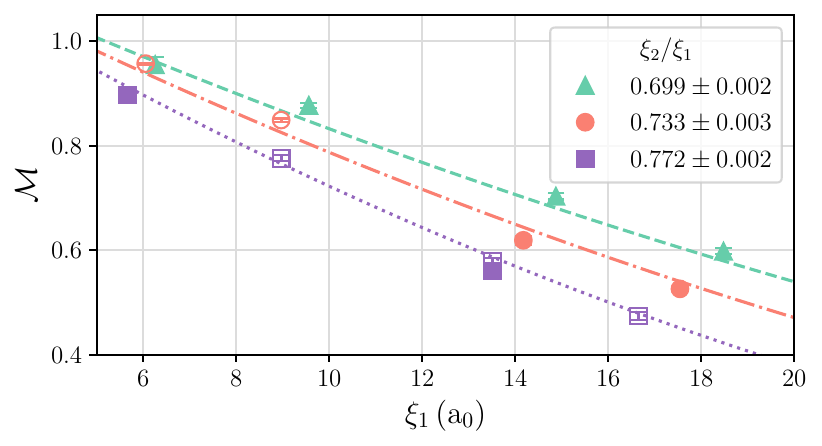}
    \caption{$\mathcal{M}$ plotted as a function of $\xi_1$ for three different values of fixed $\alpha \equiv \xi_2/\xi_1$. Since the memory can be decreased by either increasing or decreasing $\xi_1$ depending on the conditions of the experiment, this strongly indicates the presence of another length scale. The solid lines are the fits to the data using the model presented in the main text of this letter and are to help guide the eye. Closed (open) markers come from trials where the first (second) waiting time is varied. Statistical error bars are present on each data point, though they are typically smaller than the marker.}
    \label{fig:fixed_alphas}
\end{figure}

We then move onto the latter case allowed by Eq. \ref{eq:final_isotropic_result}; that this new length scale is isotropic, or $\ell_\perp = \ell_\parallel$. In this case, we would expect that memory loss does not depend on $\xi_1$, and therefore would not depend on \Tw{1} or \tw{1}. The vertical lines in Fig. \ref{fig:M_vs_alpha}  indicate instead that memory does indeed depend on both the first waiting time and temperature through $\xi_1$. To emphasize this point, in Fig. \ref{fig:fixed_alphas}, we show the values of memory \Memory for each of the fixed values of $\alpha$ from Fig. \ref{fig:M_vs_alpha}, but now plotted solely as a function of $\xi_1$. From these data, we see that memory very clearly depends on $\xi_1$, although perhaps surprisingly, it \emph{decreases} as $\xi_1$ \emph{increases}. Since we see experimentally the explicit $\xi_1$ dependence in memory, and therefore memory loss, we conclude that $\ell_\perp \neq \ell_\parallel$.

At this point, we know that both obvious cases for $V_\ell$ admitted by Eq. \ref{eq:final_isotropic_result} are consistent with our experimental data. However, we know that the length scales involved with determining the statistically independent regions of volume $V_\ell$ do not all dynamically scale as $\xi_1$, and we know they are not the same in each direction in space. The only anisotropy present in the problem is due to the ac magnetic field, though this anisotropy exists in spin-space. Regardless, we assume then that $\ell_\parallel$, the length scale coaxial with the ac field,  is the independent length whereas the lengths $\ell_\perp$  orthogonal to the ac field dynamically scale as $\xi_1$. This real-space anisotropy, when substituted into Eq. \ref{eq:final_isotropic_result}, results in 
\begin{align}
    \frac{\overline{\Delta V}}{V_{\xi_1}} &= \frac{4\pi}{3}\left(\frac{\xi_1}{\ell} \right)\alpha^3, \label{eq:final_anisotropic_result}
\end{align}
where we have taken the proportionality constant between $\ell_\perp$ and $\xi_1$  to be unity for simplicity, and have redefined $\ell_\parallel$ to be $\ell$. As is shown by our Fig. 3 in the main text, Eq. \ref{eq:final_anisotropic_result} provides data collapse among the memory measured in all 48 independent double-waiting-time experiments. This collapse strongly suggests that there is indeed another length scale $\ell$ in these experiments, but it has a real-space anisotropy in its character.

Finally, we can finish our predicted form of memory $\mathcal{M}$ for our data from Eq. \ref{eq:M_expression}. Substituting in Eq. \ref{eq:final_anisotropic_result}, we find
\begin{align}
    \mathcal{M} &= w \left( 1 - \frac{4\pi}{3\ell}\xi_1\alpha^3 \right)^{p/d},\label{eq:3D_DeltaV_over_V}
\end{align}
our functional form of memory given in the main text.  Now, the results in Fig. \ref{fig:fixed_alphas} makes sense. Our model predicts that if $\alpha$ is fixed, any increase in $\xi_1$ will lead to an increase in memory loss by Eq. \ref{eq:3D_DeltaV_over_V} due to the increasing ratio of $\xi_1/\ell$.


\bibliography{prl_bib.bib}
